\documentclass[a4paper,twocolumn,11pt,accepted=2024-03-21]{quantumarticle}
\pdfoutput=1
\usepackage[utf8]{inputenc}
\usepackage[english]{babel}
\usepackage[T1]{fontenc}
\usepackage{amsmath}
\usepackage{amssymb}
\usepackage{hyperref}

\usepackage{tikz}
\usepackage{lipsum}

\usepackage[numbers]{natbib}

\usepackage{braket}
\usepackage{bbold}

\newcommand{\e}{\mathrm{e}}
\renewcommand{\i}{\mathrm{i}}
\newcommand{\eqdef}{\overset{\text{def.}}{=}}

\newcommand{\tr}{\mathrm{tr}}
\renewcommand{\d}{\mathrm{d}}

\newcommand{\densmat}{\hat{\rho}}
\newcommand{\h}{\hat{H}}

\newcommand{\hint}{\hat{H}_{\text{int}}}
\renewcommand{\a}{\hat{a}}
\newcommand{\ad}{\hat{a}^{\dagger}}
\newcommand{\ak}{\hat{a}_k}
\newcommand{\akd}{\hat{a}_k^{\dagger}}

\newcommand{\hc}{\text{h.c.}}

\renewcommand{\Re}{\mathrm{Re}}

\begin{document}

\title{From Non-Markovian Dissipation to Spatiotemporal Control of Quantum Nanodevices}

\author{Thibaut Lacroix}
\email{thibaut.lacroix@uni-ulm.de}
\affiliation{Institut f\"ur Theoretische Physik und IQST, Albert-Einstein-Allee 11, Universit\"at Ulm, D-89081 Ulm, Germany}
\affiliation{%
 SUPA, School of Physics and Astronomy, University of St Andrews, St Andrews KY16 9SS, UK}
 \affiliation{
 Sorbonne Universit\'{e}, CNRS, Institut des NanoSciences de Paris, 4 place Jussieu, 75005 Paris, France
}%
\orcid{0000-0002-5190-040X }
\author{Brendon W. Lovett}
\affiliation{%
 SUPA, School of Physics and Astronomy, University of St Andrews, St Andrews KY16 9SS, UK}
\author{Alex W. Chin}
\affiliation{
 Sorbonne Universit\'{e}, CNRS, Institut des NanoSciences de Paris, 4 place Jussieu, 75005 Paris, France
}%
\maketitle

\begin{abstract}
Nanodevices exploiting quantum effects are critically important elements of future quantum technologies (QT), but their real-world performance is strongly limited by decoherence arising from local `environmental' interactions. 
Compounding this, as devices become more complex, i.e. contain multiple functional units, the `local' environments begin to overlap, creating the possibility of environmentally mediated decoherence phenomena on new time-and-length scales. 
Such complex and inherently non-Markovian dynamics could present a challenge for scaling up QT, but --on the other hand--  the ability of environments to transfer `signals' and energy might also enable sophisticated spatiotemporal coordination of inter-component processes, as is suggested to happen in biological nanomachines, like enzymes and photosynthetic proteins.   
Exploiting numerically exact many body methods (tensor networks) we study a fully quantum model that allows us to explore how propagating environmental dynamics can instigate and direct the evolution of spatially remote, non-interacting quantum systems. 
We demonstrate how energy dissipated into the environment can be remotely harvested to create transient excited/reactive states, and also identify how reorganisation triggered by system excitation can qualitatively and reversibly alter the `downstream' kinetics of a `functional' quantum system. 
With access to complete system-environment wave functions, we elucidate the microscopic processes underlying these phenomena, providing new insight into how they could be exploited for energy efficient quantum devices.
\end{abstract}

\section{Introduction}
The advent of the so-called \emph{second quantum revolution}\cite{macfarlane_quantum_2003,deutsch_harnessing_2020} has seen remarkable advances in the stabilization and exploitation of quantum effects in man-made nanostructures, opening the gates for an ever-growing array of technological applications which extract functional advantages from non-classical properties such as coherence and entanglement \cite{nielsen_quantum_2010, pascal_degiovanni_physique_2020, masahito_hayashi_quantum_2006,grynberg_introduction_2010, kok_introduction_2010,aspelmeyer_cavity_2014}. 
However, such quantum properties -- particularly when distributed over multiple components/qubits -- remain inherently fragile and are easily destroyed by the unavoidable interactions of the functional `system' states with the macroscopically large number of degrees of freedom that comprise their electromagnetic, electronic, or vibrational `environments' \cite{Breuer,Weiss}. 
The theory, simulation and understanding of such `open quantum systems' is thus central for the development of quantum technologies and, as we shall expose here, could even provide the insights necessary to turn system-environment interactions from a problem into a potent \emph{resource}.       

While a number of strategies have been devised to mitigate the deleterious actions of environmental phenomena such as relaxation and decoherence, the most common (practical) approach is to try to insulate (decouple) the working quantum systems from environmental fluctuations, e.g. by working at low temperatures, or using nanostructuring to suppress the relevant spectrum of environmental excitations and noise \cite{esmaielpour2021hot}, or via dynamical decoupling~\cite{viola_dynamical_1999}.                
Yet, on the other hand, emerging theoretical interest in quantum devices for harvesting and transforming energy has pointed to a number of ways in which dissipative and noisy processes could actually \emph{enhance} the efficiency of energy transfer and transduction tasks, c.f. purely classical or fully coherent operations. 
Exploiting these effects requires an optimal, and not necessarily weak, coupling between systems and environments. Notable examples of such environment-assisted phenomena include exciton and charge transfer in photosynthesis, photovoltaics and batteries \cite{Mohseni_environment-assisted_2008, Plenio_dephasing-assisted_2008, Caruso_highly_2009,wertnik2018optimizing,ghosh2021fast, quach_superabsorption_2022}, state control and energy transport in qubit networks \cite{potovcnik2018studying,maier2019environment,hansom2014environment}, and the operations of `machines' in quantum thermodynamics  \cite{kosloff2019quantum,deffner2019}. 
Indeed, there have even been proposals to use dissipative processes to implement universal quantum computation and information processing in nanoengineered environments \cite{Verstraete2009,Bermudez2013}.

\begin{figure*}
    \centering
    \includegraphics[width=\textwidth]{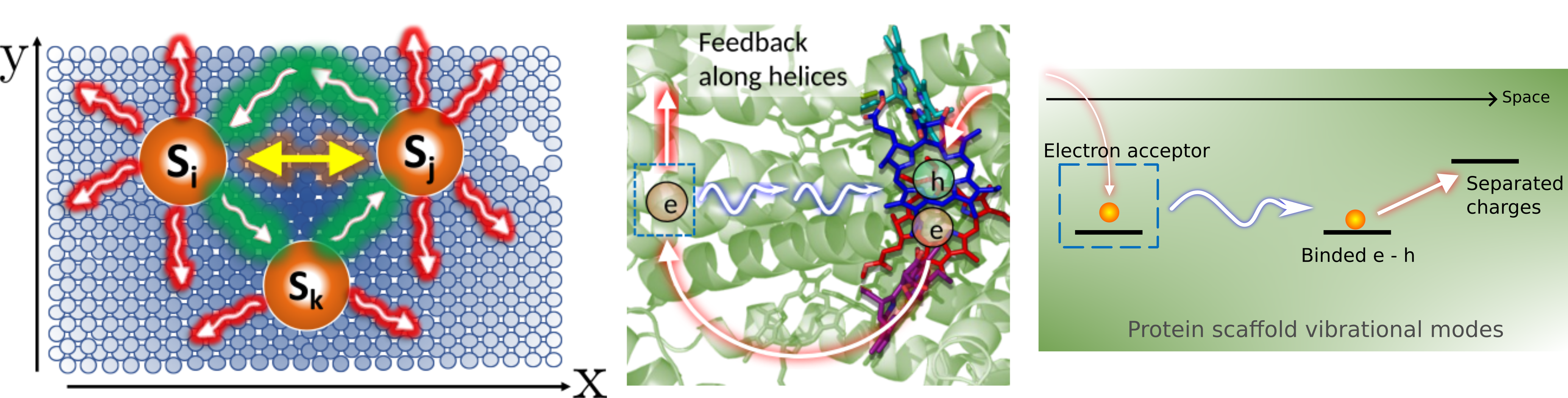}
    \caption{(a) Spatially distributed systems dissipate energy into their environment and/or cause local deformations that typically propagate and are lost in the bulk medium (red arrows). When systems are packed into nanoscale regions, a significant fraction of environmental excitations will encounter neighbouring systems and influence their dynamics (green arrows), even if the systems are uncoupled (yellow arrow indicates coherent coupling). These interactions are retarded, i.e. depend on the speed of signal propagation and the separation of the systems, providing new time and length scales to their now cooperative dissipative dynamics. Example in $1D$: (b) in photosynthetic reaction centers, pigments are held by a protein scaffold that can dissipatively mediate vibrations and structural reorganisation to coordinate exciton (eh pair) splitting, electron transfer and hole refilling in different locations (separated by $4-5$ nm) on timescales from the fs to the µs \cite{muh_nonheme_2013}. (c) The charge separation process decribed in (b) can be abstracted into an acceptor, bound exciton and charge separated states both coupled to the vibrational modes of their protein scaffold. This is the type of model-system studied in this paper.}
    \label{fig:model}
\end{figure*}

This nascent appreciation that environments and their (possibly non-Markovian) dynamics could be active and potentially programmable components of future quantum devices raises exciting possibilities that could be accessible with current nanofabrication techniques and experimental probes \cite{Groblacher2015,li2020non,liu2011experimental,khurana2019experimental,madsen2011observation}. In this work we address a relatively unexplored aspect of `environment-assisted' phenomena that is of relevance for all of the examples and topics given above. Most theoretical descriptions of interacting arrays of open quantum systems utilize models in which each component (qubits, chromophores, quantum dots, etc.) interact with `local', independent environments, i.e. while the components may interact with each other, their dissipative environments do not. Energy and/or information dissipated into these local environments is forever lost to the global multicomponent system. However, as the density of components increases -- as required for more sophisticated quantum devices -- the independence of these local environments becomes harder to justify \cite{sarovar_detecting_2020}: propagating perturbations (excitations) of their common medium  at one location become able to affect the dynamics of spatially remote systems, and maybe even do so on timescales that could be comparable to the intrinsic inter-system dynamics (see Fig.~\ref{fig:model}(a)).
Moreover, due to the retarded nature of these environmental `signals', the subsequent dynamics of each component depends on the whole history of previous system-environment interactions of \emph{every} component. The resulting complex, multiscale decoherence phenomena present a new challenge for controlling or mitigating the effects of quantum noise, however such spatiotemporal effects may also present a mechanism for collective, co-operative and non-linear feedforward/feedback responses to external perturbations, and this emerging paradigm is the focus of this article.

To predict the behaviour of such concerted, history dependent actions in a extended system under dissipation, we need to go beyond the usual assumptions of Markovian (i.e. memoryless) bath dynamics: the dynamics are manifestly non-Markovian\footnote{We would like to stress here that we use the term `non-Markovian' as a characterization that in order to obtain (or understand) the correct system's dynamics, the memory of the past interactions between the system and its environment must be taken into account~\cite{Breuer, Weiss, rivas2014quantum, li_concepts_2018}.
We do not use this term in relation to specific methods to calculate the system's density matrix.}~\cite{rivas2014quantum,de2017dynamics}, implying that they are also non-perturbative and can only be simulated with state-of-the-art numerical techniques \cite{oviedo2016phase,strathearn, Pollock,schroder2019tensor,lambert_modelling_2019, somoza_dissipation-assisted_2019, tanimura_numerically_2020, fux2021efficient, ye_constructing_2021, Cygorek2022,del2018tensor, rams_breaking_2020, de_vega_thermofield-based_2015, landi_nonequilibrium_2022, pollock_non-markovian_2018, guo_tensor-network-based_2020, white_non-markovian_2022}.
Despite the challenge, the need to explore such transient out-of-equilibrium phenomena is highlighted by physical examples in which environment (structure)-mediated communication and feedback are thought to play a key role, such as the coordination of multiscale, multielectron processes across photosynthetic proteins, electron transfer in metabolism and multistage catalysis \cite{muh_nonheme_2013, yuly_energy_2021,chaillet2020static, fourmond2019understanding,djokic2018artificial}. 
For example, in the case of reaction center proteins (Fig.~\ref{fig:model}(b)), extended $1D$ structural elements (alpha helices) couple charge-induced  mechanical/vibrational relaxation at the two functional (donor/acceptor) sites of the system, so that the presence or absence of a charge at a site can modify the activity of the other on relevant timescales \cite{muh_nonheme_2013}.
It is to be expected that environment-mediated phenomena will be strongly enhanced in low dimensional systems, and especially in $1D$, as propagating environmental excitations will necessarily encounter the other working components.

Let us consider the case given in Fig.~\ref{fig:model}~(b) of concerted charge separation in photosynthetic complexes in more detail. It can be abstracted into a model where the acceptor of a previously separated electron could be described as a TLS comprising an occupied state and an unoccupied state.
This acceptor, a stable excitonic state located at the entrance of the reaction centre (RC), and a separated electron - hole state would be coupled to the same vibrational modes of the protein scaffold.
This abstracted model is shown in Fig.~\ref{fig:model}~(c).
Of course, in order to describe completely the biological charge separation process, other states would need to be considered to describe the refilling of the hole, the transport of the electron, etc. However, our abstracted model, which we will study in detail in what follows, serves as a guide to what the full dynamics would be.

When the electron extracted from a previous exciton reaches the acceptor, i.e. the TLS switches to the occupied state, the induced change of conformation of the protein scaffold and the resulting deformation propagates (with velocity $c \sim 200\ \mathrm{m\cdot s^{-1}}$) towards the entrance of the RC ($r \sim 4\ \mathrm{nm}$ away), where a new exciton has been admitted, and reaches it a few dozens of picoseconds later.
The new mechanisms for activity at a distant site, which we will present in Sec.~\ref{sec:results} are actuated in this case by the electron being accepted onto the acceptor.  In this setting these would lead to two scenarios.
The `transient mechanism' could induce a splitting of the bound exciton in a few $\mathrm{ps}$ to the separated state with a lifetime of $\sim 10\ \mathrm{ps}$.
Alternatively, in the `static mechanism' the charge separation could happen on a timescale of the order of several hundred $\mathrm{ps}$ with a lifetime determined by the dwelling time of the previously split electron on the acceptor.
Both these timescales are compatible with the overall timescale $\sim \mathrm{\mu s}$ of the actual charge separation process.

To be clear, we do not pretend that our model is an accurate representation of the charge separation mechanism happening in photosynthetic RC.
We simply want to illustrate that the type of dissipative, environment-induced, and concerted energy tuning mechanisms that we will demonstrate in this paper is relevant to understand the dynamics and the principles of systems such as those presented in Fig.~\ref{fig:model}.
Indeed, it has been raised previously that studying the role of the environment in engineering energy levels and directing energy transport is crucial in order to understand biological systems at the microscopic level~\cite{marais_future_2018, cao_quantum_2020, kim_quantum_2021}.

Other -- somewhat arbitrarily chosen --  real-world $1D$ systems where strong environmental signalling effects could be expected include entangled triplet exciton dynamics in pi-conjugated polymers and coupled quantum dot emitters grown in nanowires~\cite{pandya2020optical,wang2019gate}.  
In general, but particularly in molecular matter, injecting `system' excitations causes the local structure to relax to a new equilibrium position, and in doing so key system properties such as energy gaps or couplings to other systems can be strongly modified in the new conformation \cite{schroder2019tensor,worth2004beyond}. 
As the establishment of this new global conformation must proceed through the propagation of local reorganization dynamics, dramatic changes at distal locations can be effected at later times \cite{leitner2008energy}. 
Such a propagating reorganization could be leveraged to act as a perturbation, a `signal', used to modify the state of a system after a given event has happened, and/or provide the energy to push remote systems out of equilibrium. 
Alternatively, the same signal could be used, not to control a quantum system, but as a non-destructive sensor of the transitions happening in a system.

To investigate such possibilities, in this paper we introduce a fully quantum mechanical model that describes how the sudden excitation of a two-level system (TLS) in a bosonic environment can induce spatiotemporal reorganization dynamics.
This dynamics is capable of \emph{reversibly} triggering and controlling the quantum dynamics of a second TLS not directly coupled to the first.
Exploiting a numerically exact tensor networks method, we demonstrate that the energy of propagating environmental `signals' can be harvested by the second TLS to `activate' (populate) metastable excited states that could trigger downstream processes with timing and lifetime information that could be leveraged to match them to other processes.      
Strikingly, when the TLSs are placed closer to one another, analysis of the complete system-bath wave functions -- a powerful capability of our tensor-based approach -- reveals that static, non-local reorganization can stabilize this activated state, leading to 100$\%$ quantum yield and a lifetime that is only limited by the lifetime of the first TLS excited state. 
We further demonstrate that these `transient' (energy harvesting) and `conformational' (thermodynamic) physics are robust at finite temperatures. We suggest that the rich phenomenology of this remote control OQS model, together with the forensic capabilities of our new simulation methods, could open up new design concepts for multi-component quantum `machines' and functional materials.

\section{Model: Quantum switch}
\label{sec:derivation}
\begin{figure*}
    \centering
    \includegraphics[width=\textwidth]{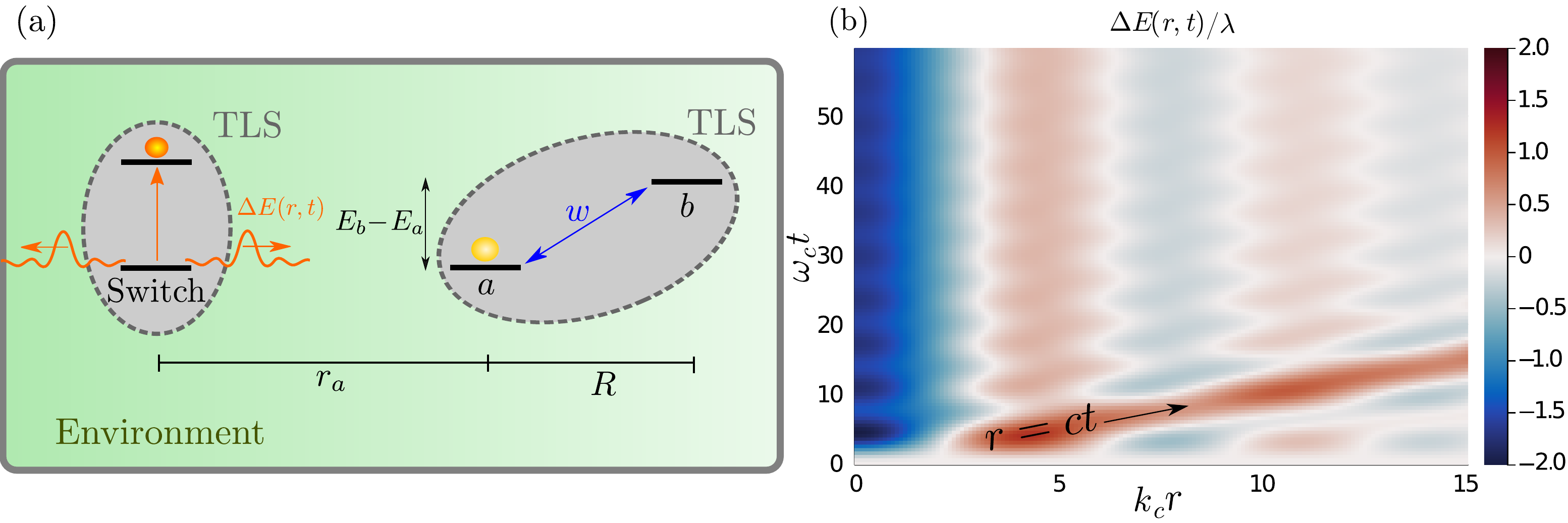}
    \caption{(a) Representation of the system. A TLS made of sites $a$ and $b$, whose energy difference is $E_b - E_a$, is interacting with the same bosonic environment as the so called ``switch" site excited at $t = 0$. This excitation will transiently perturb the environment which will influence the energy levels of the two sites. (b) Heatmap of the energy shift $\Delta E(r, t)$ for an Ohmic spectral density with a hard cutoff induced by an excitation at $r=0$ with $\lambda$ the bath reorganization energy, $c = 1$ and $k_c = 1$. The energy shift is composed of a static stabilizing contribution centered on the excitation and two destabilizing contributions (for convenience only $r>0$ is shown), whose amplitudes are half the size, propagating away from the excitation at the speed of sound $c$.}
    \label{fig:schematic_system}
\end{figure*}
Figure \ref{fig:schematic_system}(a) shows a schematic representation of the model we shall investigate, which consists of two uncoupled and spatially separated TLSs that couple to the excitations of a common environment. As discussed earlier, such a model is an abstraction of charge separation in photosynthesis. However, it is also 
inspired by other 1D mechanically deformable structures coupled to qubits or optoelectronic excitations -- e.g. nanomechanical resonators \cite{arcizet_single_2011, yeo_strain-mediated_2014, treutlein_hybrid_2014},or organic polymers \cite{kohler_electronic_2015}. Our environment is taken to be a 1D \emph{continuum} of spatially extended harmonic oscillators, as would be expected in an extended mechanical structure, and could naturally represent phononic, vibrational, photonic, or magnon baths in specific applications. 
For simplicity, we take these environment modes to be dispersionless 1D plane waves.
The first part of the system is a single site that can be either occupied or unoccupied that we call \emph{`the switch'} ($S$).
The second part of the system, composed of two sites $a(b)$ at a distance $r_a (r_b)$ from the switch, is a TLS with an energy gap $E_b - E_a$ and coherent tunneling coupling $w$.
The Hamiltonian describing the system, the environment and their interaction is
\begin{align}
    \label{eq:Hamiltonian}
   \h  = &\sum_{i= a,b}E_i \hat{P}_i + w\left(\ket{a}\bra{b}+\hc\right)+ \int_{\mathbb{R}}\d k~ \hbar\omega_k\akd\ak \nonumber\\ &+ \sum_{i=S,a,b}\varsigma_i\hat{P}_i \int_{\mathbb{R}}\d k~ \left(g_k^i\e^{\i k r_i}\ak + \hc\right)\ ,
\end{align}
where $\varsigma_i = \pm 1$ determines the sign of the coupling, $\hat{P}_i$ projects onto the localised system state $\ket{i}$ in which site $i$ is excited, and $\akd$ is the bath creation operator for the plane-wave mode of energy $\hbar\omega_k = \hbar c |k|$ with $k$ the wave-vector of the mode and $c$ the speed of sound. The coupling coefficients between the site $i$ and the mode $k$ are $g_k^i\e^{\i k r_i}$. 
This coupling induces energy shifts of the system sites that are proportional to the linear displacement of the environmental modes, a coupling that is frequently found in open system literature and applications in physics, chemistry and biology.
Sites $a$ and $b$ can formally be either in an occupied state or an unoccupied state.
However, due to the excitation preserving nature of their interaction, the only two accessible states are $a$ occupied and $b$ unoccupied (which we call $\ket{a}$) and $b$ occupied and $a$ unoccupied (which we call $\ket{b}$), hence this is formally a TLS.
The sites $a$ and $b$ have the same coupling coefficients $g_k^{a/b} = g_k$ and the ratio of the switch coupling coefficients and the site coupling coefficients is $g_k^S/g_k = \sqrt{\kappa}$.
The coupling coefficients define the bath spectral density $J(\omega) = \sum_k |g_k|^2 \delta (\omega-\omega_k)$, which we take to have the widely used Ohmic form $J(\omega) = 2\alpha \omega H(\omega_c - \omega)$ where $\alpha$ is the (dimensionless) strength of the system-bath coupling, $H(x)$ is the Heaviside function and $\omega_c=c|k_c|$ is the cut-off frequency corresponding to the largest wave-vector $k_c$ of the bath.
We use $\omega_c$ as our reference energy scale in all numerical calculations. 

\section{Methods}

\subsection{Numerical Calculations}
To perform numerically exact simulations of the dynamics that result from the Hamiltonian in Eq.~(\ref{eq:Hamiltonian}), we use a tensor networks method that relies on the mapping of normal modes environments coupled linearly to a system onto a 1D semi-infinite chain.
Doing so enables us to write the joint state of the system and the bath as a \emph{Matrix Product State} (MPS).
This method is called \emph{Thermalised - Time Evolving Density matrix with Orthonormal Polynomials Algorithm} (T-TEDOPA) \cite{Chin2010, Tamascelli2019, Lacroix2021}.
More details about T-TEDOPA and the tensor networks used can be found in Appendix~\ref{sec:tn}.
The maximal number of bath modes considered was $\sim 400$ and the maximal bond dimension of the MPS Ansatz was $\chi \sim 30$.
For a discussion of the size of the bond dimension see Appendix~\ref{sec:convergence}.
The time-evolution was performed using an implementation of the one-site Time-Dependent Variational Principle (1TDVP) for MPS~\cite{haegeman_time-dependent_2011, haegeman_unifying_2016, paeckel_time-evolution_2019, MPSDynamics}.
The 1TDVP method preserves the unitarity of the time evolution and conserves the total energy.
This method delivers the entire joint wave-function $\ket{\psi(t)}$ of the system and environment (and its equivalent at finite temperature), and thus allows us to analyse the full quantum dynamics of propagating reorganization including system-environment entanglement and environment-induced dissipation.

\subsection{Initial States}\label{sec:init}
We are interested in studying the impact of the perturbation of the environment induced by the switch being excited on the dynamics of the remote TLS made of sites $a$ and $b$.
To do so, we wish to remove the environmental dynamics associated with the relaxation of the bath being `quenched' by an excitation on site $a$.
Hence, we consider at $t = 0^{-}$ the steady state of the \{System + Environment\} resulting from the relaxation of the bath in presence of an excitation on site $a$.
This steady state is the product state of an excitation on site $a$ and a displaced vacuum state.
The initial joint state for the simulation at $T=0$ is thus
\begin{align}
    \ket{\psi(t=0^+)} &= \ket{1,1,0}_S\otimes\hat{D}(\{\delta_k\})\ket{\{0\}}_B
\end{align}
where $\ket{1,1,0}_S$ is the system's state with an excitation on the switch site, an excitation on site $a$ and no excitation on site $b$, $\ket{\{0\}}_B$ is the vacuum state of the bath and $\hat{D}(\{\delta_k\})$ is a multi-mode displacement operator.
In App.~\ref{appendix:displaced} we explain how $\hat{D}(\{\delta_k\})$ is described in our tensor networks framework.
The displacement comes from allowing the bath to reach equilibrium with the system's state $\ket{0,1,0}_S$ where there are no excitation on the switch site and an excitation on site $a$ which is positively coupled to the environment ($\varsigma_a = 1$).
The relaxed initial state of the environment is a coherent state where each mode $k$ has been displaced by an amount $\delta_k = -\frac{g_k}{\omega_k}\e^{-\i k r_a}$.
Then, at $t=0^+$, the switch site is excited.
At finite temperature, the initial system state is unchanged but the environment is described by an undisplaced Gibbs state at inverse temperature $\beta = (k_B T)^{-1}$
\begin{align}
    \hat{\rho}(t=0^+) = \ket{1,1,0}_S\bra{1,1,0}_S\otimes\exp\left(-\beta\h_B\right)/Z\ ,
\end{align}
where $\h_B = \int_{\mathbb{R}}\d k~ \hbar\omega_k\akd\ak$ is the free bath Hamiltonian, and $Z$ is the bath partition function.

\section{Results\label{sec:results}}
\subsection{Reorganization Dynamics}
We first present some physical intuition for the action of the reorganization dynamics.
We consider the case where the switch is excited, and look at the energy shift it would induce at a remote site at a position $r_\gamma$, if it were occupied. 
This is an exactly soluble model since it does not take account of the dynamics of any of the remote sites.  
The bath part of the interaction Hamiltonian (the last term in Eq.~(\ref{eq:Hamiltonian})) can then be treated as an effective external field that creates a space-dependent shift of the system energies given by
\begin{align}
    \Delta E(r_{\gamma}, t) = & \varsigma_{\gamma}\tr\left[ \int_{\mathbb{R}}g_k(\ak\e^{\i k r_\gamma} + \hc)\d k\densmat_B(t)\right].
    \label{eq:work}
\end{align}
$\Delta E(r_{\gamma},t)$ can be interpreted as work performed by the environment on site $\gamma$ due to its displacement at position $r_\gamma$ \cite{deffner2019}.
Injecting an excitation onto the switch site ($r=0$) at $t=0$ would create a time-dependent energy shift of a system at $r_{\gamma}$, given by
\begin{align}
    \Delta E (r_{\gamma}, t) = &\frac{-\varsigma_{\gamma}2\lambda\sin(k_c r_{\gamma})}{k_c r_{\gamma}}\nonumber\\
    &+ \varsigma_{\gamma}\sum_{\xi=\pm 1}\frac{\lambda\sin(k_c(r_{\gamma} - \xi ct))}{k_c(r_{\gamma} - \xi ct)},
    \label{eq:shift_hard}
\end{align}
where $\lambda = \int_0 ^\infty J(\omega)\omega^{-1}d\omega = 4\alpha\omega_c$ is the bath reorganization energy.
More information on the bath reorganization energy and the derivation of Eq.~(\ref{eq:shift_hard}) can be found in App. \ref{appendix:reorganization} \& \ref{appendix:derivation} respectively.
Figure \ref{fig:schematic_system}(b) shows the time evolution described by Eq.~(\ref{eq:shift_hard}) for $\varsigma_{\gamma}=1$; the local relaxation of the environment stabilizes the switch excitation on a timescale $\approx \omega_c^{-1}$ (negative energy shift), while two outgoing waves with positive amplitude propagate away from the origin with velocity $c$. 
These constitute the signals that will act on the distant TLS in the following sections. 
The hard cutoff in $J(\omega)$, modulates $ \Delta E (r, t)$ at the wavelength $2\pi/k_c$ due to the Gibbs phenomenon. We note that the positive propagating shifts are a manifestation of conservation of energy under the mechanical distortion of the medium induced by the presence of an excitation on the switch site: in the local bath approximation these contributions are assumed to propagate away rapidly without encountering any other system components, i.e. their energy is essentially lost and the local relaxation is \emph{irreversible} (See Fig. \ref{fig:model}a).    
At long times only the static contribution centered at the position of the excitation remains (first term of Eq.~(\ref{eq:shift_hard})), although it has a non-negligible spatial extension ($\approx k_c ^{-1} $) and acts like a potential energy shift on the other sites. 
Additionally, this potential also shows spatial oscillations that will also play an important role in the full quantum dynamics, below. 

\subsection{Remote Transient Activation}
\begin{figure*}
    \centering
    \includegraphics[width=\textwidth]{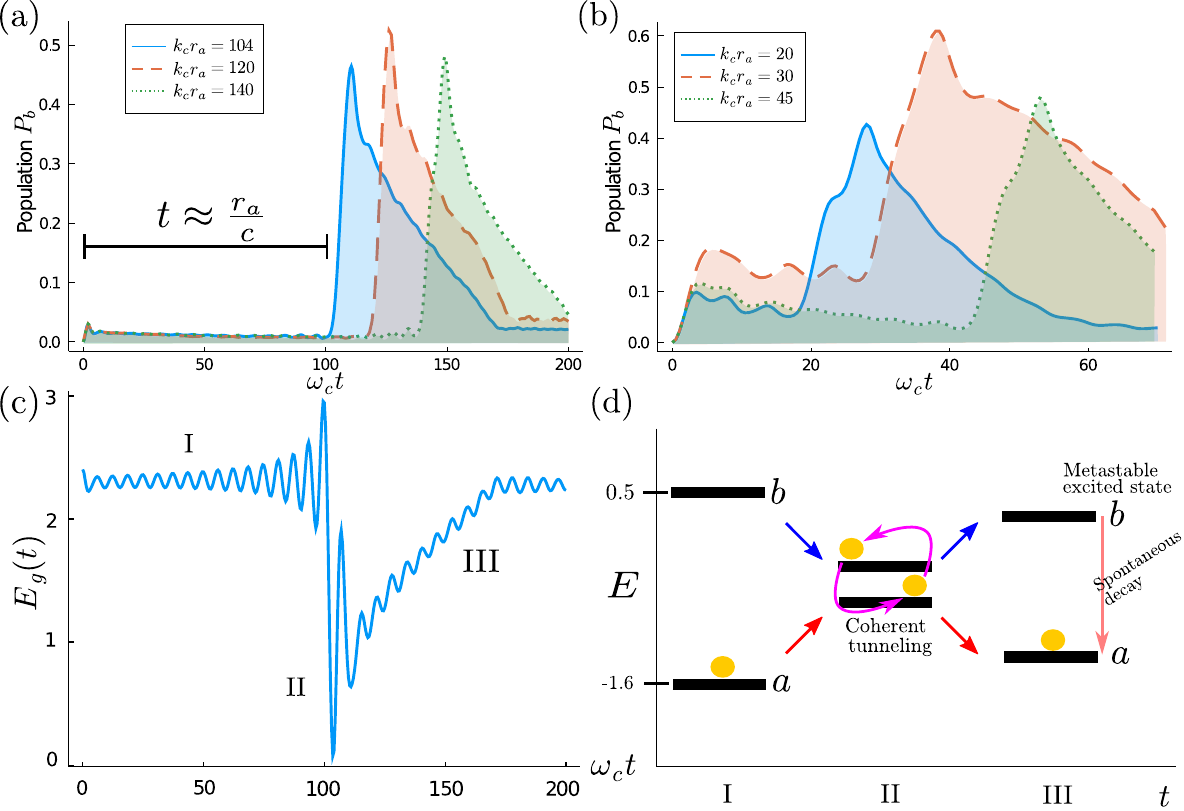}
    \caption{Site $b$ population for several distances $r_a$ between the switch and site $a$ with $\alpha = 0.2$, $\kappa\alpha = 1.2$, $w =  0.15\omega_c$. (a) $T = 0$, there is no dynamics until $t=r_a/c$. The separation between the sites $a$ and $b$ is $k_c R = 4$. (b) Finite Temperature $\beta = 10$ presents similar results. (c) Effective energy gap $E_g$ between the two sites as a function of time for $k_c r_a = 104$ at $T=0$. When the energy perturbation reaches the site it momentarily closes the gap. Three regions are labelled: I before the perturbation reaches the sites; II when the perturbation is closing the gap; and III after the passage of the perturbation. (d) Schematics of the relative spacing between the sites $a$ and $b$ energy levels.}
    \label{fig:LongdistanceR=4}
\end{figure*}

To study the influence of the transient energy perturbation on the distant TLS at zero temperature, we consider an initial state in which site~$a$ is occupied, and positively coupled to the environment ($\varsigma_a = 1$), but both site~$b$ and the switch are unoccupied.
We allow the bath to reach equilibrium with this system state configuration as explained in Sec.~\ref{sec:init}.
Throughout, we take $E_b-E_a = 0.5\omega_c$, $w=0.15\omega_c$ and $\alpha=0.2$, so that including the reorganization energy due to the initial relaxation, the total energy gap of the TLS is $E_b - E_a + 2\lambda = 2.1\omega_c$. 
This is over ten times larger than the coupling $w$, so there is negligible population dynamics in the absence of signals from the switch. 
At $t=0^+$, we inject an excitation onto the switch site, hence triggering the reorganization signal. 
Figures~\ref{fig:LongdistanceR=4}(a)-(b) show the dynamics of the population of site $b$ for different switch-site distances $r_a$ and fixed $k_c R=4$ ($R=r_b-r_a$) both at zero ($\beta = \infty$) and finite ($\beta = 10$) temperatures.
All other parameters are held fixed, the coupling signs set to $\varsigma_S = \varsigma_a = - \varsigma_b = 1$ and $\kappa\alpha=1.2$.
The impact of interaction signs on the dynamics is discussed in App.~\ref{appendix:sign}. 
The choice of opposite signs for sites $a$ and $b$ has been made here to enhance the possibility of a population transfer.\\

When the perturbation generated by the switch-induced reorganization reaches the system at $t\approx r_a/c$, a sudden transfer of population is initiated that can pump over $50\%$ into the higher-energy site $b$. 
Once the perturbation has passed, this population decays back to site $a$ because of downhill energy relaxation. 
This transfer of population occurs both at $\beta = \infty$ and $\beta \neq \infty$.
Figure \ref{fig:LongdistanceR=4}(c) shows the evolution at zero temperature of the TLS energy gap as it dramatically closes, as the perturbation raises the energy of site $a$, and Fig.~\ref{fig:LongdistanceR=4}(d) is a schematic drawing of the spacing of the sites' energy levels at different moments in time.
This transient near-resonance allows coherent transfer of population through an adiabatic transition, creating a metastable excited state whose energy could, for example, be directed towards a desired function. 
Zero and finite temperatures present the same qualitative dynamics where relaxation of the bath induced by the switch causes a sudden large transfer of population into site $b$.
To reduce computation time and illustrate the general nature of the phenomenon, the finite temperature results shown in Fig.~\ref{fig:LongdistanceR=4}(b) are for smaller $k_cr_a$ than the zero temperature results.
At finite temperature the initial state we use assumes a bath that has not relaxed when there is an excitation on site $a$; this initial relaxation can be seen in the additional smaller transient population transfer starting at $t=0^+$ in Fig.~\ref{fig:LongdistanceR=4}(b).
This initial relaxation is of no particular physical significance as it is at least an order of magnitude lower than switch-induced transient population transfer that can be observed later on in the dynamics.
With an unoccupied switch, this initial transfer of population at finite temperature decays back to zero on a timescale of $\omega_c t \sim 40$.
The gain of population when the perturbation reaches site $a$ happens at a similar rate in both cases.
The available energy of this state has come directly from the work performed on the system by the reorganization dynamics, and optimizing this would be an interesting area for further work. 
We note a difference in the decay rates between the zero and finite temperatures cases.
The highly non-exponential decay of the excited state for zero temperature in Fig.~\ref{fig:LongdistanceR=4}(a) is suggestive of non-Markovian dissipation.
The timing of the population transfer event can be controlled via distance or propagation speed.
Indeed, modification of the spectral function may provide a way to tune the excited state lifetime to match its downstream function. 
Additionally, the philosophy of our model is conserved if one swaps the roles of the switch site and the remote TLS.
If one considers the switch site as the system of interest, then the TLS can be seen as a sensor which can monitor transitions of the switch without direct interactions.
In that case Figs.~\ref{fig:LongdistanceR=4}(a)-(b) can be reinterpreted as an indirect way to access the state of the switch site by `listening' to the environment.

\subsection{Remote Permanent Activation}
\begin{figure}
    \centering
    \includegraphics[width=\columnwidth]{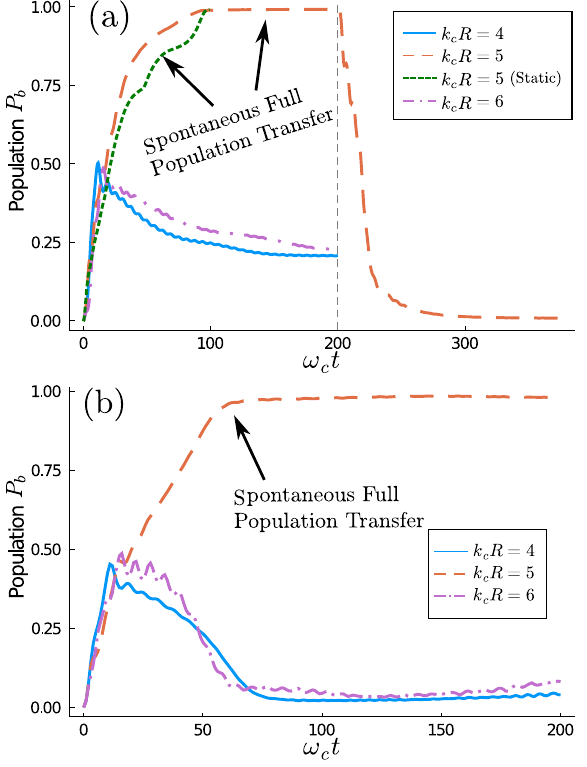}
    \caption{(a) For $k_c r_a = k_c R = 5$ there is a full population transfer at $\beta = \infty$ both when the full reorganization dynamics is taken into account and when only the static part of the landscape is considered. In the former case, the population decays back to site $a$ when the switch site is de-excited at $\omega_c t = 200$ (dashed vertical line and orange background). (b) Finite Temperature $\beta = 10$ also presents a full population transfer.}
    \label{fig:permanent}
\end{figure}

\begin{figure*}
    \centering
    \includegraphics[width=\textwidth]{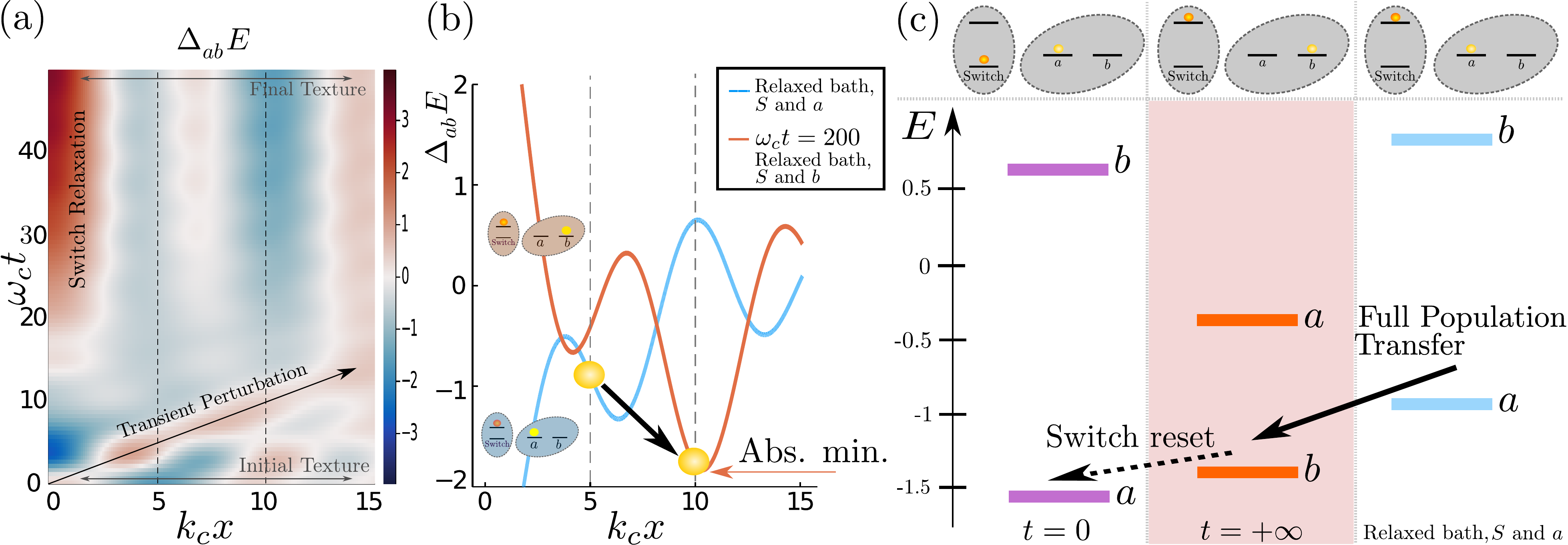}
    \caption{(a) Energy landscape perceived by sites $a$ and $b$ (marked by dashed lines) as a function of time and space for $k_c R = 5$, $\alpha = 0.2$, $\kappa\alpha = 1.2$, and $\beta = \infty$. (b) Energy shifts for the state a long time after the transient perturbation has passed (i.e. relaxed environment in the presence of the excited switch and site $b$) and the state where the environment reached an equilibrium when the switch is excited and the population of the remote TLS is localized on site $a$ ($w = 0$). As the absolute minimum of the energy landscape moves from the location of site $a$ to that of site $b$, the population evolves correspondingly. For convenience in (a) and (b) only the positive half of space is shown. (c) Schematics of the absolute energy levels of sites $a$ and $b$ for the different configurations.}
    \label{fig:trajectories_full}
\end{figure*}

A striking change of behavior is observed when the TLS is brought closer to the switch site, causing the fate of the excited state to become highly sensitive to the switch-TLS distance. 
Figure~\ref{fig:permanent}(a) shows the TLS dynamics of the system at zero temperature  for smaller $r_a$, where for $k_c r_a = k_c R = 5$, the population is \emph{permanently} transferred from site $a$ to site $b$ with $100\%$ yield even after the propagating perturbation passes. 
For $k_c r_a = k_c R = 4,6$ the yield drops to less than $25\%$. 
An analogous behavior can be observed at finite temperature (see Fig.~\ref{fig:permanent}(b)), where at $\beta = 10$ for $k_c r_a = k_c R = 4,6$ a transient transfer of population to the higher-energy state on site $b$ is induced, whereas for $k_c R = 5$ a full transfer is achieved.
However the kinetics of the full population transfer are different.
Contrary to the $\beta = \infty$ case, at finite temperature the $k_c r_a = k_c R = 4,6$ populations decay to the lower energy eigenstate.
This is due to the presence of thermal fluctuations that allow the remnant population on site $b$ to explore the energy landscape and decay back to the lower energy site $a$.
The role of the coupling strength $\alpha$ on the yield of the population transfer is investigated in App.~\ref{appendix:strength}.
The stability of population transfer at $k_c R = 5$ suggests a lasting thermodynamic change in the energy ordering of the TLS sites, which we have traced to the role of the static shift $\Delta E(r,t\rightarrow \infty)$ induced by the switch (Eq.~(\ref{eq:shift_hard})). 
This can be directly visualised by computing $\Delta_{ab} E(x,t) =  \sum_{i=a,b}\bra{\psi(t)}\varsigma_{i}\hat{P}_{i} \int_{\mathbb{R}}\d k~ \left(g_k\e^{\i k x}\ak + \hc\right)\ket{\psi(t)}$.
This gives a representation of the evolution of the transient `energy landscape' perceived by the TLS, and is composed of a large number of two-time two-point correlation functions that can be readily evaluated within our many-body tensor networks simulation framework. 
To gain more insight into these effects Fig.~\ref{fig:trajectories_full}$(a)$ shows a heatmap of $\Delta_{ab} E$ for the full population transfer case ($k_c R = 5$) at $\beta = \infty$. 
At $t=0^+$ the environment starts to relax because of the presence of the switch and generates a static negative energy shift at $x = 0$ and two propagating positive energy shifts (only the positive $x$ one is shown) which propagate at the speed of sound. 
When the right-propagating transient perturbation reaches site $a$, it lifts its energy level and thus initiates population transfer to site $b$. 
However, once the perturbation passes, the population transfer continues.
Fig.~\ref{fig:trajectories_full}(b) shows a cross section of the heat maps at the final time and the energy landscape in the hypothetical case where the environment has relaxed to an equilibrium state in presence of an excited switch and a populated site $a$.
It can be seen that after the switch is fully relaxed, an occupied site $b$ now corresponds to the global ground state of the system, and local environmental dissipation will thus drive the system to this state, as illustrated in Fig.~\ref{fig:trajectories_full}(c).
\subsection{Energy Landscape Texture}
Light can be shed on the distance-dependence of this stabilized transfer by also considering the \emph{`texture'} of the energy shift induced by the $\mathrm{sinc}$-shaped static part of the switch-generated energy perturbation (see Fig.~\ref{fig:schematic_system}(b)).
The switch generates a static negative energy shift of large amplitude around the origin and several local minima and maxima that alternate in sign with a wavelength $2\pi/k_c$. 
The further away from the switch, the lower the amplitude of these extrema (at $k_c x = 10$, $90\%$ of the amplitude has been lost already). 
In the $k_c R=5$ case, site $a$ is close to a maximum of this landscape and is raised in energy. Site $b$ sits on the second maximum, but because it couples to the environment with a negative sign, this static contribution corresponds to a stabilizing shift. 
The energy landscape texture depends on the nature of the bath spectral density.
Sharp decays or super-Ohmicity for example can be a source of texture.
For the parameters used in our simulations, these static shifts cause the bare energy of site $b$ to become \emph{lower} than site $a$, after the switch has relaxed. 
Small changes $\delta R / R < 1$ in the position of the sites would not drastically alter the full population transfer results as the important factor for this process to happen is that the static contribution is enough to invert the energy gap.
Although the initial relaxation of the environment around site $a$ still creates a large barrier for population transfer, the static switch potential now renders the population on site $a$ metastable. 
Thus, the activation of the switch has primed a permanent conformational change in the environment that drives full population transfer. 
This is directly evidenced in Fig. \ref{fig:permanent}(a), showing that an excitation initially localized on site $a$ undergoes spontaneous and complete transfer to site $b$, when the environment is initially relaxed in the presence of a populated switch. 
This observation has a strong implication, namely that this transition should be \emph{reversible} once the switch is depopulated, as is verified in Fig.~\ref{fig:permanent}(a) for $k_c R = 5$ when we remove the switch excitation at $\omega_ct = 200$.\\
All these transitions can be summarized by looking at the sites' energy levels for different configurations, as shown in Fig.~\ref{fig:trajectories_full}(c).

Reversibility of transitions is of paramount importance, from a control perspective, as it enables the switching of the state of the TLS using a mechanical medium; our model highlights how spatial variations in the non-local energy shift could be exploited for dissipatively `locking in' activated states.
This reversible 100\%-yield population transfer also enables the possibility of information transduction from, for instance, an electromagnetic medium (a laser-pulse used to excite the switch site) to a mechanical medium.
An additional way to frame this effect of spatial variation of the reorganization energy landscape is to consider it as a simple form of environment engineering.
The excited switch site precisely located could be considered as part of the environment of the TLS --- which would effectively be described with a new spectral density --- whose dissipative dynamics now spontaneously populates the TLS site $b$.
Hence, one could design a dissipative energy landscape by combining several switch-like systems and placing them at selected positions in space.
Finally, in contrast to dynamically activated metastable states, we note that `permanently activated' states are favoured thermodynamically, i.e. do not harvest energy from the triggering of the switch directly; the work done by the environment has gone into creating the \emph{driving force} for the \emph{dissipative} population transfer. 
The deterministic movement of, say, charged excitations to new sites could be used to trigger a wide range of chemical and mechanical processes which could provide catalytic coordination or sensing functions.

\section{Conclusion}
Our open quantum model of non-Markovian, environmentally mediated signalling, allows us to leverage unavoidable dissipative processes to reversibly control the transition of a remote TLS.
Thanks to advanced simulation methods, two distinct classes of controlled process --- transient and permanent activations --- have been identified both at zero and finite temperatures, and further work will explore the efficiency of these mechanisms in single-shot and cyclic, i.e. engine, operation. 
Crucially our quantum switch model provides a natural platform to explore strongly non-classical effects in remote signalling, and could naturally be extended to multiparticle systems to explore spatiotemporal entanglement dynamics driven by dissipative processes. 
Indeed, this quantum switch model could provide a reversible and non-invasive means of controlling, entangling and probing qubits.
Physical 1D platforms such as cold atoms, superconducting qubit chains and ion traps could also be used as quantum simulators for such physics. 
The study of common environment dissipative physics that can give rise to non-local and non-Markovian effects is crucial in the context of quantum engineering in order to be able to channel dissipation toward useful functions, and to mitigate induced cross-talk errors in multi-component nanodevices.
The possibility, opened by our model, for the environment to be in a superposition of two different displacements of opposite magnitudes hints toward indefinite causal order of dissipative processes \cite{Chiribella2013, oreshkov_quantum_2012}.
In another direction, our model could be refined in multiple ways to describe realistic (bio-)chemical systems.
For instance, our present theoretical tools could be used to approach structured spectral densities found by \textit{ab initio} methods \cite{renger2012normal, Dunnett2021}, anharmonic effects \cite{morgan2016nonlinear,leitner2001vibrational}, or Hamiltonian topologies that account for more complex connectivity between systems and/or environments \cite{schroder2019tensor}.
Another perspective for this model is to provide a new set of processes to analyse the intricate physical effects happening in biological systems made of complexes of organic molecules such as allostery \cite{Changeux2011, hilser_structural_2012, fourmond2019understanding, liu_allostery_2016} or multi-electron processes \cite{yuly_energy_2021}.

\acknowledgments

TL, AWC and BWL thank the Defence Science and Technology Laboratory (Dstl) and Direction G\'en\'erale de l’Armement (DGA) for support through the Anglo-French PhD scheme. 
BWL acknowledges support from EPSRC grant EP/T014032/1.
All numerical results were produced using our open source Julia software package \texttt{MPSDynamics.jl}, which is freely available at \href{https://github.com/shareloqs/MPSDynamics}{https://github.com/shareloqs/MPSDynamics}. 

\bibliographystyle{quantum}
%

\onecolumn
\appendix

\section{Tensor Networks Simulations, chain mapping}\label{sec:tn}
\begin{figure}[h]
    \centering
    \includegraphics[width=\columnwidth]{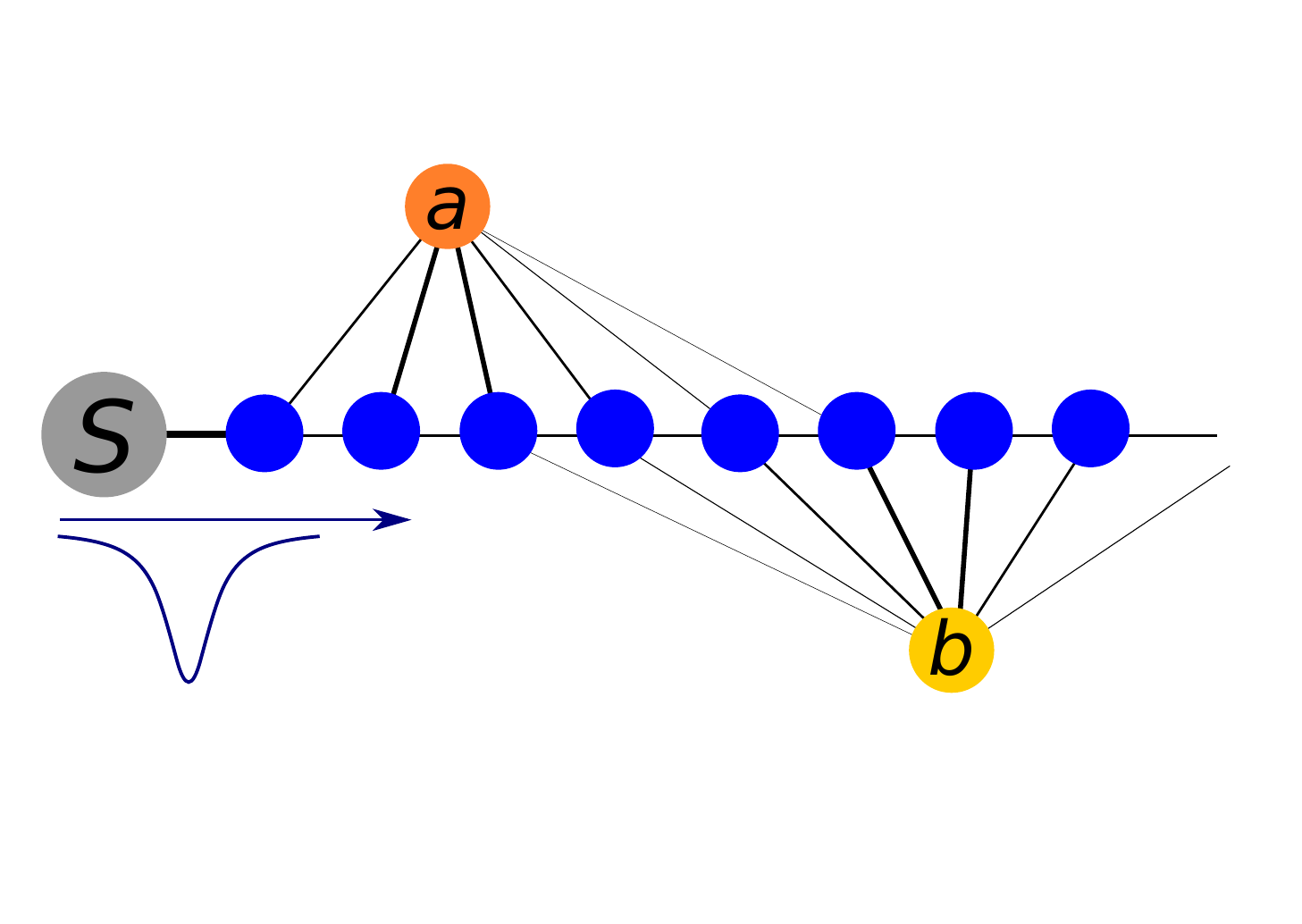}
    \caption{Schematic representation of the mapping between the system and one of the two chains representing the environment. The switch is only coupled to the first chain mode and sites $a$ and $b$ are coupled to every mode with a coupling that is maximal around $n \propto r_{a/b}$.}
    \label{fig:chainmapped}
\end{figure}

In order to study the population dynamics of the energy barrier, we use a numerically exact method able to handle non-Markovian dynamics that performs the time-evolution of the system and the environment together (enabling us to access not only the system's dynamics but also the environment's dynamics).
To do so, the non-interacting continuous environment is mapped to two tight-binding $1d$-chains with long-range couplings between the system and the environment for the propagating ($k>0$) and counter-propagating ($k<0$) modes \cite{Chin2010, Lacroix2021} using a unitary transformation $U_n(k)$.  
The resulting bath Hamiltonian is 
\begin{align}
     \h_E &= \sum_{n} \omega_n (\hat{c}^\dagger_n \hat{c}_n + \hat{d}^\dagger_n \hat{d}_n) + t_n(\hat{c}^\dagger_n \hat{c}_{n+1} + \hat{c}^\dagger_{n+1} \hat{c}_n + \hat{d}^\dagger_n \hat{d}_{n+1} + \hat{d}^\dagger_{n+1} \hat{d}_n)\ ,
\end{align}
where $n$ labels the discrete modes of the new chains, $\hat{c}_n^{\dagger}$ and $\hat{d}_n^\dagger$ are the creation operators of these chains with onsite energies $\omega_n$ and hopping energies $t_n$, defined by introducing unitary transformations
\begin{align}
     \a_{k\geq 0} &= \sum_{n=0}^{+\infty} U_n(k) \hat{c}_n\ , \label{eq:UnitaryTransform1}\\
     \hat{a}_{-k\geq 0} &= \sum_{m=0}^{+\infty} V_m(k) \hat{d}_m\ ,
     \label{eq:UnitaryTransform2}
\end{align}
with the matrix elements
\begin{align}
    U_n(k) = V_n(k) = g_k P_n(k)
\end{align}
where $\{P_n\}_{n \in \mathbb{N}}$ are orthonormal polynomials with respect to the measure $\mu(k) = |g_k|^2 \eqdef J(k)$ such that $P_0(k) = 1$ and
\begin{align}
    \int_{0}^{+k_c} P_n(k) P_m(k) J(k)\d k = \delta_{n,m}\ .
    \label{eq:orthogonality}
\end{align}
The nature of the polynomials thus depends on the spectral density of the bath.
We chose an Ohmic spectral density with a hard cut-off (here at $k_c$) $J(k) = 2\alpha c^2 k H(k_c - k)$, where $\alpha$ is a coupling strength and $H$ the Heaviside step function.
In that case, the $P_n$ are Jacobi polynomials.
Another useful property of these polynomials is that they obey a recurrence relation
\begin{align}
    P_n(k) &= (k - A_{n-1})P_{n-1}(k) + B_{n-1}P_{n-2}(k)\ ,
    \label{eq:recurrence}
\end{align}
where $A_n$ is related to the first moment of $P_n$ and $B_n$ to the norms of $P_n$ and $P_{n-1}$ \cite{Chin2010}.
This recurrence relation can be used to construct the polynomials with the conditions that $P_0(k) = 1$ and $P_{-1}(k) = 0$, and the recurrence coefficients are used to define the onsite energy $\omega_n$ and the hopping energy $t_n$.
The interaction Hamiltonian transforms as
\begin{align}
    \hint &= \sum_{i}\varsigma_i\hat{P}_{i}\sum_{n=0}^{+\infty} \Big(\gamma_n(r_{i})(\hat{c}_n + \hat{d}^\dagger_{n}) +\hc \Big)\ ,
\end{align}
where $\varsigma_i = \pm 1$, as defined in Eq.~(1), and the new coupling coefficients are
\begin{align}
    \gamma_n(r_{i}) = \int_{0}^{+k_c} g_k\e^{\i k r_{i}}U_n(k)\d k\ .
\end{align}
The different system sites couple differently to the chain modes and for a given site the coupling strength across modes is not uniform.
A system site at a position $r$ couples mostly to a specific region of the chain around the mode $n \propto r$.
Since the switch site is situated at the origin, it couples solely to the first mode of each chain (which is a linear combination of \emph{all} the normal modes of the environment).\\
The joint state of the system and the environment can be described with a Matrix Product State representation and the Hamiltonian can we written in the form of a Matrix Product Operator \cite{Lacroix2021}.\\
In practice, the chains cannot be kept semi-infinite and need to be truncated at a given chain mode $N_m$.
The number of environmental modes that has to be considered is set by the total simulation time (here $\omega_ct_\text{max} = 200$).
An excitation created in the chain-mapped environment should not have the time to reach the truncated end of the chain and be reflected toward the system: this would lead to unphysical recurrences induced by the finite nature of the environment after truncation.
For the simulation time considered here, 400 modes fulfil this condition.\\
We use a tensor network implementation of the time-dependent variational principle to perform the time-evolution \cite{MPSDynamics}.
A schematic representation of how the system couples to the chain-mapped environment is shown in Fig.~\ref{fig:chainmapped}.

\section{Convergence and bond dimension\label{sec:convergence}}
\begin{figure}[h]
    \centering
    \includegraphics[width=0.7\columnwidth]{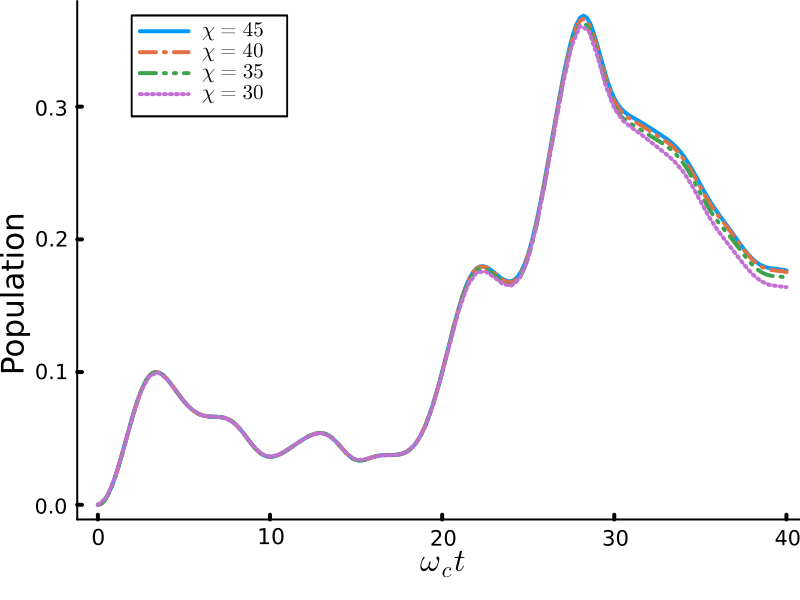}
    \caption{Convergence plot for the population of site $b$ for different values of the bond dimension $\chi$ with the same parameters as in Fig.~\ref{fig:LongdistanceR=4}~(a) and $k_c r_a = 20$. The simulation results are already nearly converged at $\chi = 30$ with negligible deviations from the converged results $\chi = 45$.}
    \label{fig:convergence-transient}
\end{figure}

The size of the bond dimension $\chi$ is an important convergence parameter for tensor network simulations.
This parameter encodes the amount of entanglement between bi-partitions of the quantum state.
For usual one-site implementation of the TDVP algorithm, the bond dimension is constant and uniform along the MPS.
Figure~\ref{fig:convergence-transient} shows the population of site~$b$ obtained for several values $\chi$ of the bond dimension from $\chi = 30$ to $\chi = 45$ where the simulation results are converged.
We can see that for times $t \lesssim \frac{r_a}{c}$ the dynamics are already converged for $\chi = 30$.
After the perturbation induced by the switch has reached the TLS, an extra amount of entanglement is needed leading to a slightly larger value of the required bond dimension.
This increase of the bond dimension for $t \geq \frac{r_a}{c}$ is a sign of the creation of entanglement between the switch and the TLS mediated by the environment.
We note that the qualitative behavior of the population is the same for $\chi = 30$ and $\chi = 45$ and the maximal quantitative deviation is of the order of $6\%$.

\begin{figure}[h]
    \centering
    \includegraphics[width=0.7\columnwidth]{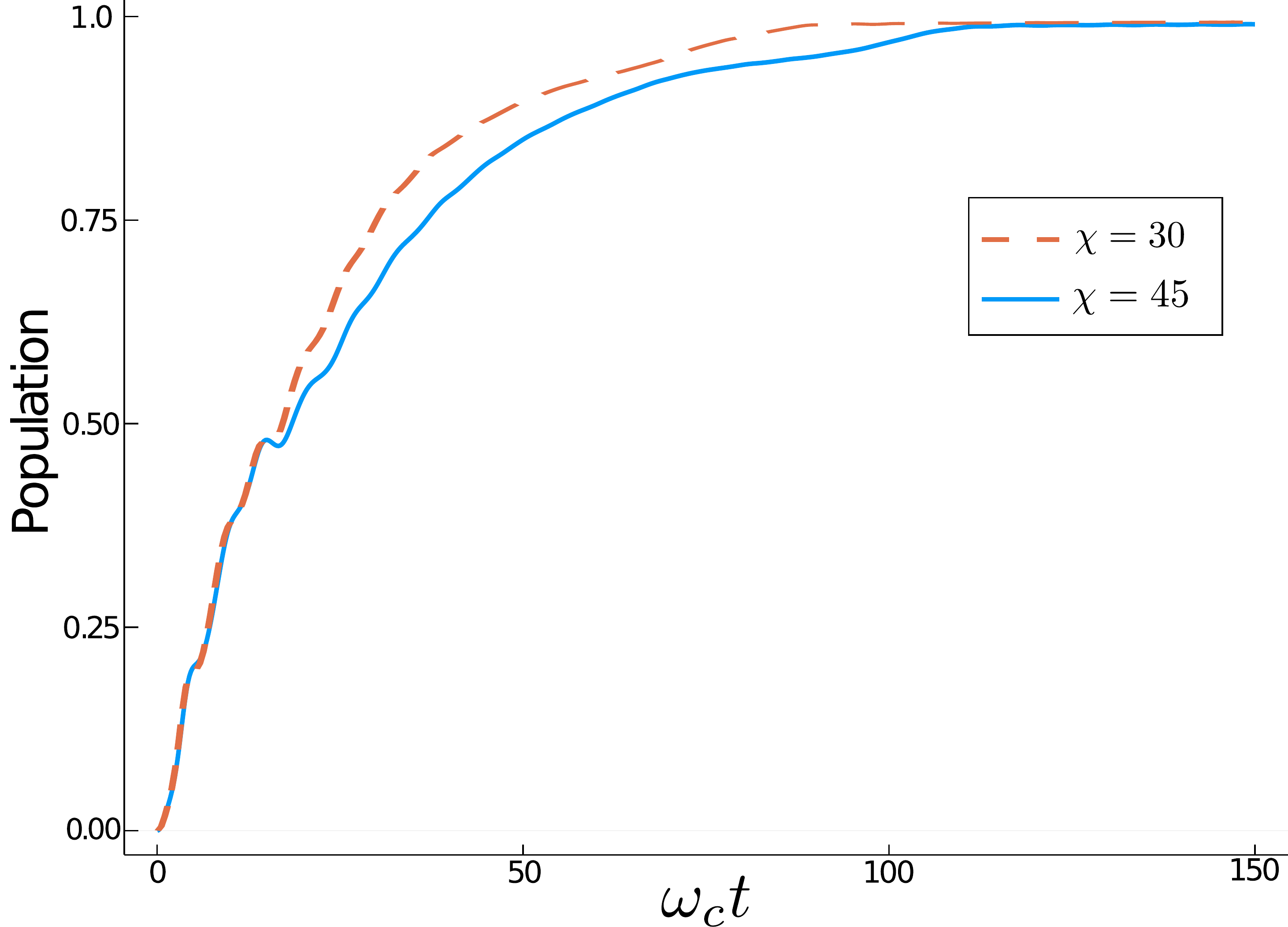}
    \caption{Comparison of site~$b$ population for $k_c r_a = k_c R = 5$ with bond dimensions $\chi = 30$ and $\chi =45$. Both cases give the same full population transfer final state in spite of deviations during transient dynamics.}
    \label{fig:convergence-permanent}
\end{figure}

Figure~\ref{fig:convergence-permanent} shows site~$b$ full population transfer when $k_c r_a = k_c R = 5$ at zero-temperature obtained with a converged bond dimension $\chi = 45$ and the results presented in Fig.~\ref{fig:permanent}.
We can see a good qualitative agreement between both results and especially that full population transfer is reached in both cases with the same timescale.

\section{Displaced bath in the chain representation}\label{appendix:displaced}
Instead of running longer simulations to let the system and the environment reach a steady state in the presence of an excitation at site $a$, we directly consider a state with a displaced bath.
One of the advantage of this is that such a state only needs to take into account of the static part of the reorganization energy.
Hence, fewer chain modes are needed because we discard the initial positive perturbations.
The complex amplitudes of the displacement $\delta_k$ of $k$-modes for a static excitation since $t=-\infty$ are 
\begin{align}
    \delta_k &= -g_k\e^{-\i k r_a}\mathrm{P}\left(\frac{1}{\omega_k}\right)\ ,
\end{align}
where $\mathrm{P}(\cdot)$ is the principal value distribution.
For any practical purpose because we use an Ohmic spectral density (which are linear in $k$) we can substitute $g_k\mathrm{P}\left(\frac{1}{\omega_k}\right)$ by $g_k/\omega_k$ if we keep in mind that $\delta_{k=0} = 0$.
The bath displacement operator $\hat{D}(\{\delta_k\})$ can be written in the chain representation
\begin{align}
    \hat{D}(\{\delta_k\}) &= \exp\left(\int_{\mathbb{R}}(\delta_k \akd - \delta_k^* \ak) \d k\right)\\
    &= \exp\Big(\int_{\mathbb{R}^+}(\delta_k \akd - \delta_k^* \ak) \d k + \int_{\mathbb{R}^+}(\delta_{-k} \ad_{-k} - \delta_{-k}^* \a_{-k}) \d k\Big)\\
    &= \exp\left(\int_{\mathbb{R}^+}(\delta_k \akd - \delta_k^* \ak) \d k\right) \times \exp\left(\int_{\mathbb{R}^+}(\delta_{k}^* \ad_{-k} - \delta_{k} \a_{-k}) \d k\right)\\
    &= \prod_n \exp\left(\int_{\mathbb{R}^+} \delta_k U_n(k)\hat{c}_n^\dagger \d k - \hc\right) \times \prod_n \exp\left(\int_{\mathbb{R}^+} \delta_k^* U_n(k)\hat{d}_n^\dagger \d k - \hc\right)\\
    &= \prod_n \exp\left(\delta_n\hat{c}_n^\dagger - \hc \right)\prod_n \exp\left(\delta_n^*\hat{d}_n^\dagger - \hc \right)\label{eq:displacementMPO}\\
    \Rightarrow\hat{D}(\{\delta_k\}) & = \hat{D}(\{\delta_n\})
\end{align}
where $\hat{D}(\{\delta_n\})$ is a displacement operator on the chain and $\hat{b}^\dagger_n$ ($\hat{c}^\dagger_n$) are the chain creation operator originating from positive (negative) $k$-vectors.
The complex displacements of chain modes are 
\begin{align}
    \delta_n &= \int_{\mathbb{R}^+} \delta_k U_n(k)\d k\\
    &= \int_{\mathbb{R}^+}\frac{-g_k}{\omega_k}\e^{-\i k r_a} g_k P_n(k) \d k\\
    \delta_n (r_a) &= -\int_{\mathbb{R}^+} \frac{J(k)}{\omega_k}P_n(k)\e^{-\i k r_a}\d k = -2\pi\mathcal{F}\left[\frac{J}{\omega}P_n\right](r_a)\ .
\end{align}
The complex displacement amplitude of a chain mode $n$ is proportional to the Fourier transform (because $J(k)$ is non-zero only for positive $k$) of the reorganization energy multiplied by the $n$th order polynomial defining the chain mapping.
For an Ohmic spectral density the displacement is proportional to the Fourier transform of the appropriate polynomial, which might be computed analytically.
For an Ohmic spectral density, if $r_a=0$ only the first site of the chain is displaced thanks to the orthogonality of the polynomials.
The chain displacement operator $\hat{D}(\vec{\delta})$ naturally has a MPO representation with bond dimension 1 as can be seen in Eq.~(\ref{eq:displacementMPO}).
\section{Sign of the interaction Hamiltonian}\label{appendix:sign}
\begin{figure}[h]
    \centering
    \includegraphics[width=\columnwidth]{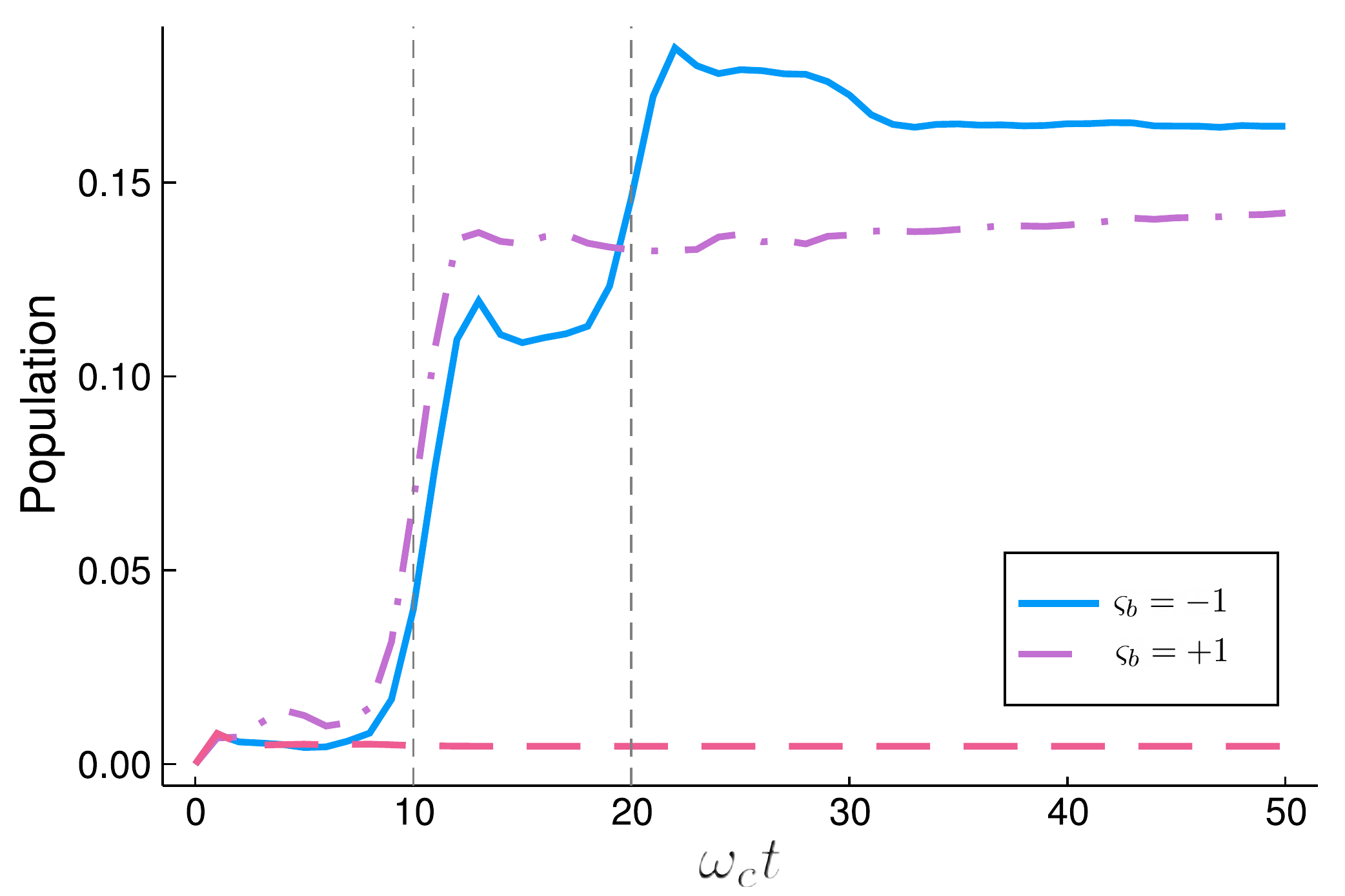}
    \caption{Population of site $b$. Vertical dashed lines are times of the peak of the perturbation reaching site $a$ and site $b$. We can see step-like population growth corresponding to the perturbation. The pink dashed line shows the absence of evolution of the population in the absence of the transient perturbation.}
    \label{fig:pop_steps}
\end{figure}

To illustrate the influence of the transient energy perturbation on the dynamics of the system, let us consider the following choice of parameters: the gap is $E_b - E_a = 0.5\omega_c$, the tunneling energy is $w =  0.15\omega_c$, the separation between the sites is $k_c R = 10$, the coupling strength is $\alpha = 0.4$, the speed of sound is $c=1$, $\kappa = 3$ and $\varsigma_a = 1$.
With such parameters the initial renormalized energy gap is $E_b - (E_a -8\alpha\omega_c) = 3.7\omega_c \gg w$, hence we expect no dynamics for the populations.
However, if the switch site is excited at $t = 0$, the transient energy perturbation should affect the energy of site $a$  and site $b$ respectively around the times $t = R/c$ and $t = 2R/c$, and could induce some dynamics because the renormalized gap is reduced for a short amount of time.
The dynamics of the gap between the two sites will also depend on the sign of $\varsigma_b$.
A positive sign would mean that the perturbation increases the gap and a negative sign reduces the gap.
Figure~\ref{fig:pop_steps} shows the evolution of the population of site $b$ for both values of $\varsigma_b$.
In both cases the dynamics is radically altered as the population grows instead of staying at zero.
When $\varsigma_b = -1$, two steps are visible and can be clearly associated with the transient energy perturbation.
In the opposite case, the population stays constant after the first step.
Indeed, the gap becoming larger when the perturbation reaches site $b$ prevents further population transfer.
With this first example we can already see that non-trivial population transfer can be induced by the environment.

\section{Definition of the reorganization energy with a toy example}\label{appendix:reorganization}
Consider a system interacting with a single harmonic oscillator of angular frequency $\omega$ with the following Hamiltonian
\begin{align}
    \h &= \h_S + \frac{\hbar\omega}{2}(\hat{X}^2 + \hat{P}^2) + g\hat{S}\hat{X}
\end{align}
where $\hat{X}$ and $\hat{P}$ are respectively the dimensionless displacement and dimensionless momentum operators of the harmonic oscillator, $\hat{S}$ is a diagonal system operator, and $g$ is the coupling strength between the system and the oscillator.
The linear coupling term between the system and the oscillator can be removed by changing $\hat{X} \to \tilde{X} = \hat{X} + \hat{A}$, \emph{i.e.} by changing the equilibrium position of the oscillator.
Hence the harmonic oscillator's Hamiltonian becomes
\begin{align}
    \h_B &= \frac{\hbar\omega}{2}(\hat{X}^2 + \hat{P}^2)\\
    &= \frac{\hbar\omega}{2}((\tilde{X} - \hat{A})^2 + \hat{P}^2)\\
    &= \frac{\hbar\omega}{2}(\tilde{X}^2 + \hat{A}^2 - 2\tilde{X}\hat{A} + \hat{P}^2)\\
    \h_B &= \frac{\hbar\omega}{2}(\tilde{X}^2 + \hat{P}^2) + \frac{\hbar\omega}{2}(\hat{A}^2 - 2\tilde{X}\hat{A})\ ,
\end{align}
and the interaction Hamiltonian becomes
\begin{align}
    \hint &= g\hat{X}\hat{S} = g\tilde{X}\hat{S} - g\hat{A}\hat{S} \ .
\end{align}
We choose $\hat{A} = \frac{g}{\hbar\omega}\hat{S}$ such that the new linear coupling term $g\tilde{X}\hat{S}$ is canceled out by the linear term $-\hbar\omega\tilde{X}\hat{A}$ from the oscillator's Hamiltonian.

The Hamiltonian has now the following form 
\begin{align}
    \h &= \h_S + \frac{\hbar\omega}{2}(\tilde{X}^2 + \hat{P}^2) + \frac{\hbar\omega}{2}\hat{A}^2 - g\hat{A}\hat{S}\\
    &= \h_S + \frac{\hbar\omega}{2}(\tilde{X}^2 + \hat{P}^2) + \frac{g^2}{2\hbar\omega}\hat{S}^2 - \frac{g^2}{\hbar\omega}\hat{S}^2\\
    \h &= \h_S + \frac{\hbar\omega}{2}(\tilde{X}^2 + \hat{P}^2) - \frac{g^2}{2\hbar\omega}\hat{S}^2 \ .
\end{align}
We define the reorganization energy $\lambda \eqdef \frac{g^2}{2\hbar\omega}$.
This reorganization energy comes from a positive contribution from the oscillator's Hamiltonian and a negative contribution twice as large from the interaction Hamiltonian.
Hence, at equilibrium ($\langle\tilde{X}\rangle = 0$), the expectation value of the interaction Hamiltonian is $\langle\hint\rangle = -2\lambda\langle S^2\rangle$.

\section{Derivation of the energy shift}\label{appendix:derivation}
From Eq.~(\ref{eq:work}), we have an analytic expression for the energy shift (with $\varsigma = +1$ for simplicity)
\begin{align}
    \Delta E(r, t) &= \int_{\mathbb{R}}2g_k\Re\left[\langle\ak\rangle_B(t)\e^{\i k r}\right]\d k\ ,
    \label{eq:energyshift}
\end{align}
and thanks to Ehrenfest theorem given the Hamiltonian in Eq.~(1)
\begin{align}
    \langle\ak\rangle_B(t) 
    &=\frac{g_k}{\omega_k}\sum_\alpha\e^{-\i k r_\alpha}\langle\hat{P}_\alpha\rangle(\e^{-\i\omega_k t} - 1)\nonumber\\&\text{~~~ assuming~~}\langle\hat{P}_\alpha\rangle(\tau) \simeq \mathrm{cst}\ .
\end{align}
Under this assumption, the energy shift can be expressed as
\begin{align}
    \Delta E(r, t) &= \sum_\gamma \langle\hat{P}_\gamma\rangle \int_{\mathbb{R}} 2\frac{J(k)}{\omega_k}\Big[\cos\big(\omega_k t -k(r - r_\gamma) \big)  - \cos\big(k(r - r_\gamma)\big)  \Big]\d k \ .
\end{align}
This energy shift can be interpreted for each system site as a wave-packet propagating away from the site with an envelope $\propto J(k)/\omega_k$, plus a static term centered on the site.
It can be readily noticed that, because of energy conservation, the energy shift summed over all space at a given time is zero.
Trivially, we can see that at $t = 0$ the energy shift vanishes.
For a single localised excitation at $r_\gamma = 0$ and for any time $t$
\begin{align}
    \int_{\mathbb{R}} \Delta E(r, t) \d r &= \int_{\mathbb{R}}\d r \int_{\mathbb{R}}\d k 2\frac{J(\omega_k)}{\omega_k}\Big(\cos\big(\omega_k t -kr \big) - \cos(kr)\Big)\ ,
\end{align}
but $\int_{\mathbb{R}}\left(\cos(\omega_k t - kr) - \cos(kr)\right)\d r = \delta(k) - \delta(k) = 0$ as $\omega_{k=0} = 0$.
Hence, conservation of energy is satisfied.
The shape of the wave-packet depends on the bath spectral density $J(k)$.
The family of Ohmic spectral densities is given by 
\begin{align}
    J(k) &= 2\alpha c^2\frac{|k|^s}{k_c^{s-1}}f_{k_c}(k)
\end{align}
where $\alpha$ is the strength of the system-bath coupling, $c$ is the speed of sound, $k_c$ is the wave-number cut-off of the bath, and $f_{k_c}$ is a cut-off function equal to $e^{-\frac{|k|}{k_c}}$ for soft exponential cut-off and to the Heaviside step function $H(k_c - k)$ for a hard one.
The parameter $s$ labels the family such that $0 < s <1$ are sub-Ohmic densities, $s = 1$ is a Ohmic density and $s > 1$ are super-Ohmic densities.

\begin{figure}
    \centering
    \includegraphics[width=\columnwidth]{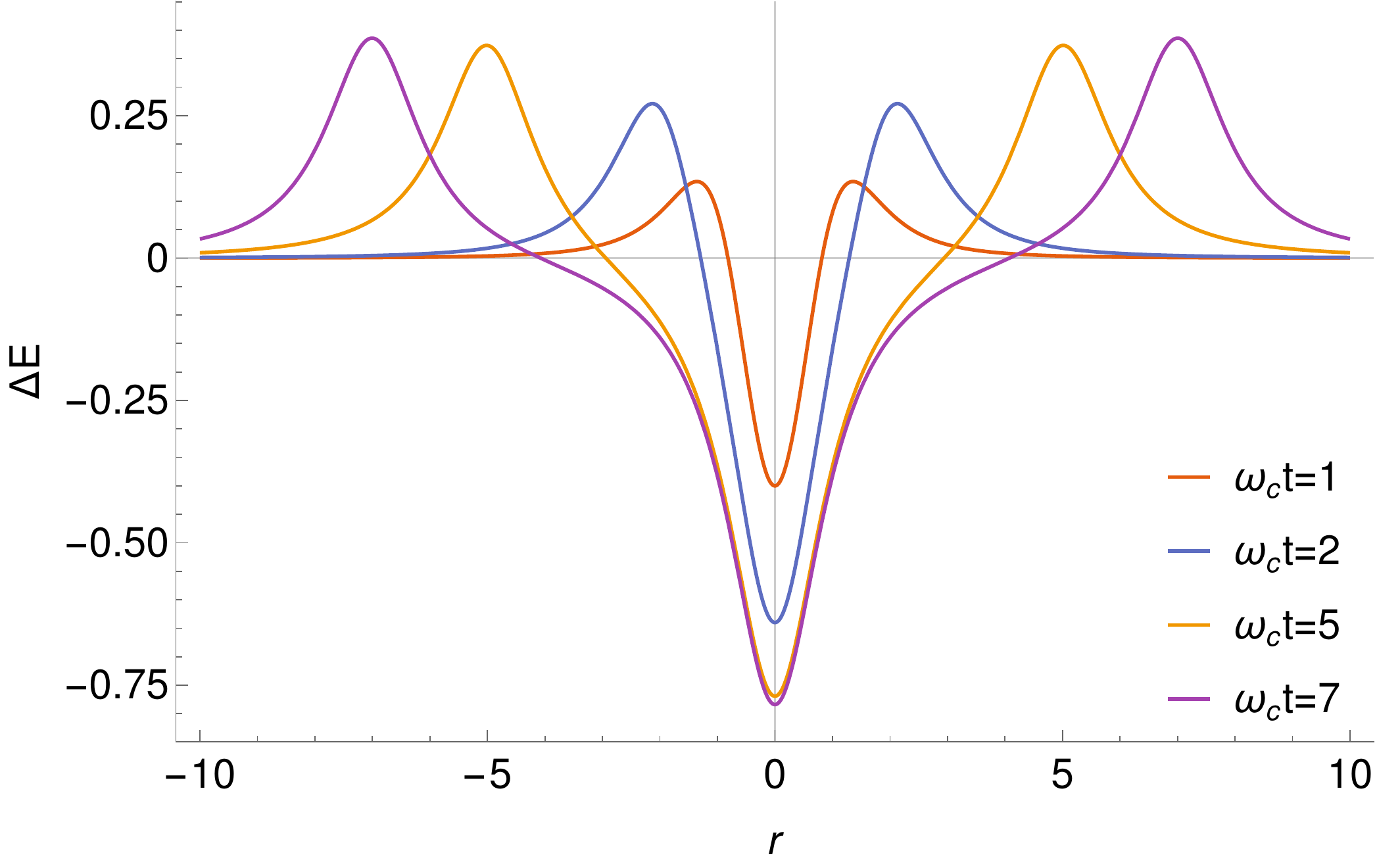}
    \caption{Energy shift $\Delta E(r, t)$ at several times for a Ohmic spectral density with a soft cutoff induced by an excitation at $r=0$ with $\alpha = 0.1$, $c = 1$ and $k_c = 1$.}
    \label{fig:soft}
\end{figure}

For a soft cutoff the energy shift is
\begin{align}
\Delta E (r,t) &= 4\alpha c k_c \Gamma(s)\Bigg( \frac{-2\cos\big(\arctan(k_c r)s\big)}{(1 + (k_c r)^2)^{s/2}} + \frac{\cos\big(\arctan(k_c (r - ct))s\big)}{(1 + (k_c(r - ct))^2)^{s/2}} + \frac{\cos\big(\arctan(k_c (r + ct))s\big)}{(1 + (k_c (r + ct))^2)^{s/2}} \Bigg) \ ,
\end{align}
which reduces in the Ohmic ($s=1$) case to 
\begin{align}
    \Delta E(r, t) &=\lambda \Bigg( \frac{-2}{1 + (k_c r)^2} + \frac{1}{1 + (k_c(r - ct))^2} + \frac{1}{1 + (k_c(r + ct))^2} \Bigg)\ ,
    \label{eq:OhmicSoft}
\end{align}
where $\lambda = 4\alpha\omega_c$ is the bath reorganization energy.
The three terms are Lorentz distributions of heights $-2\lambda$, $\lambda$ and $\lambda$ respectively and width $1/k_c$.
Figure \ref{fig:soft} shows the energy shift given by Eq.~(\ref{eq:OhmicSoft}) at different times.
The first term is the static negative contribution corresponding to twice the reorganization energy and the two other terms are destabilizing (positive) perturbations propagating away from the site.
The stationary state is realized by taking $t\to\infty$ and corresponds only to the negative static contribution around the position of the site, \emph{i.e.} twice the reorganization energy.

For the hard cut-off case, the results are qualitatively the same as in the soft cutoff case but with modulation of the wave packet at the wavelength $2\pi/k_c$ due to the Gibbs phenomenon.
The energy shift becomes
\begin{align}
    \Delta E (r, t) & = \lambda\Bigg( \frac{-2\sin(k_c r)}{k_c r} + \frac{\sin(k_c(r - ct))}{k_c(r - ct)} + \frac{\sin(k_c(r + ct))}{k_c(r + ct)} \Bigg)\ .
\end{align}

\section{Influence of the coupling strength}\label{appendix:strength}
\begin{figure}[h]
    \centering
    \includegraphics[width=\columnwidth]{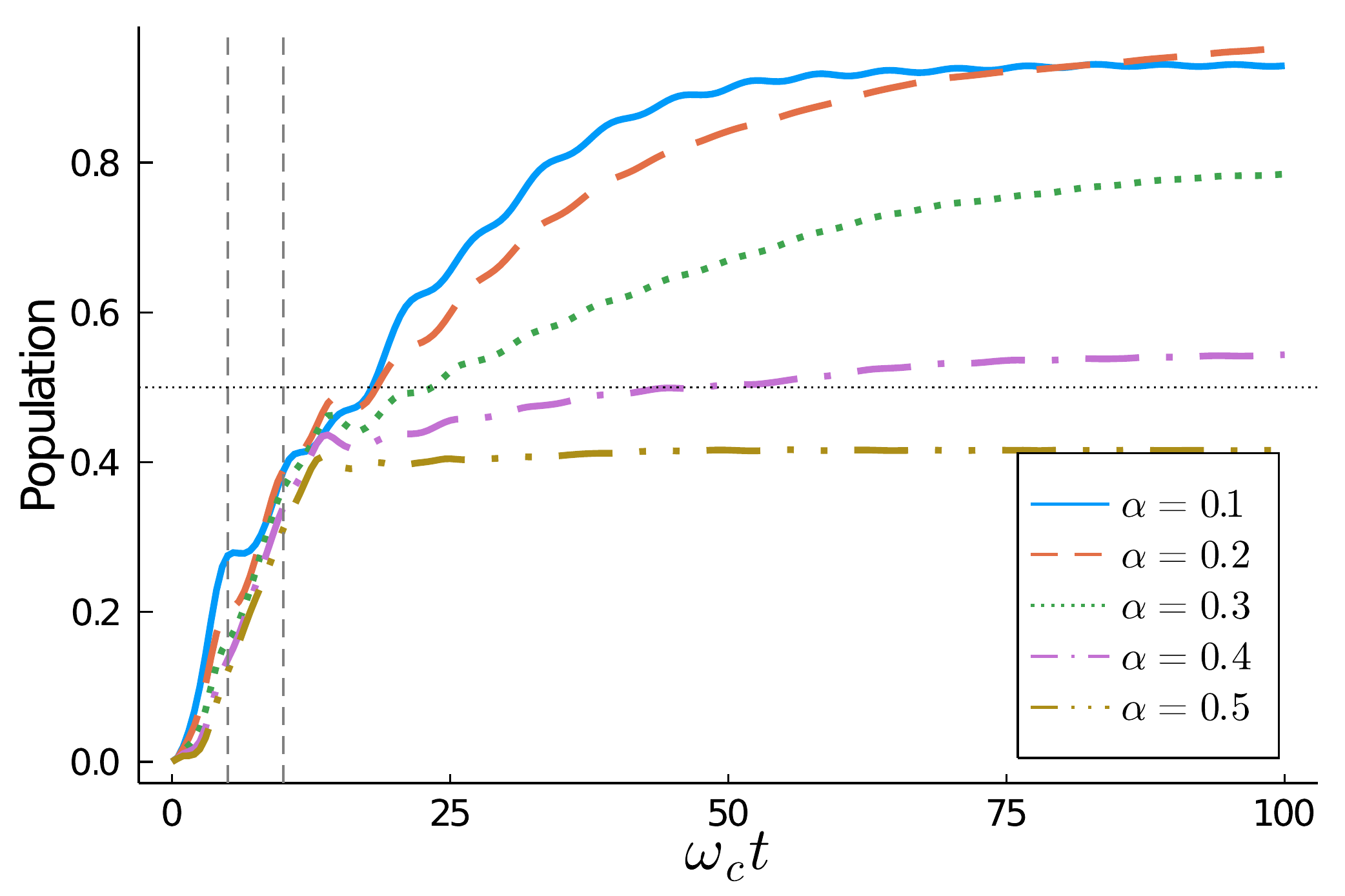}
    \caption{Population of site $b$ for different coupling strength $\alpha$ with $R =5$, $w =  0.15$, $\omega_c = 1$, $c = 1$ and $\kappa\alpha = 1.2$. The dashed vertical lines indicate the time at which the perturbation reaches the sites $a$ and $b$. The horizontal dotted line shows half population. Long lasting partial and total population transfer are initiated by the effect of the switch.}
    \label{fig:pop_transfer_alpha}
\end{figure}

The amount of population transferred can be controlled via the coupling strength between the system and the bath as it depends on the renormalized gap between the two sites.
The higher the coupling to the bath, the larger the reorganization energy.
Thus at a fixed amplitude of the energy perturbation -- holding $\kappa\alpha$ constant --, increasing the coupling strength between the system and the bath leads to a decrease of the transferred population as the barrier to be crossed becomes higher.
For larger values of $\alpha$, the occupied site $b$ states becomes even more favoured thermodynamically but, because of the increase of the initial gap, the transfer of population slows down, and can even plateau at a partial population transfer.
Figure \ref{fig:pop_transfer_alpha} shows how the population of site $b$ is impacted by $\alpha$.

\end{document}